\documentclass[aps,prd,preprintnumbers,groupedaddress,nofootinbib,amssymb,notitlepage,eqsecnum]{revtex4-1}
\usepackage{amsmath,amsthm,amssymb}
\usepackage{bm}
\usepackage{color}

\usepackage{amsfonts}
\usepackage{dcolumn}
\usepackage{hyperref}
\allowdisplaybreaks[1]
\usepackage{stackengine}

\begin{document}
\preprint{YITP-23-62, IPMU23-0013}

\title{
Instability of scalarized compact objects in Einstein-scalar-Gauss-Bonnet theories
}

\author{Masato Minamitsuji}
\affiliation{Centro de Astrof\'{\i}sica e Gravita\c c\~ao  - CENTRA, Departamento de F\'{\i}sica, Instituto Superior T\'ecnico - IST, Universidade de Lisboa - UL, Avenida Rovisco Pais 1, 1049-001 Lisboa, Portugal}

\author{Shinji Mukohyama}
\affiliation{Center for Gravitational Physics and Quantum Information, Yukawa Institute for Theoretical Physics, Kyoto University, 606-8502, Kyoto, Japan}
\affiliation{Kavli Institute for the Physics and Mathematics of the Universe (WPI), The University of Tokyo Institutes for Advanced Study, The University of Tokyo, Kashiwa, Chiba 277-8583, Japan}

\begin{abstract}
We investigate the linear stability of scalarized black holes (BHs) and neutron stars (NSs) in the Einstein-scalar-Gauss-Bonnet (GB) theories against the odd- and even-parity perturbations including the higher multipole modes. We show that the angular propagation speeds in the even-parity perturbations in the $\ell \to \infty$ limit, with $\ell$ being the angular multipole moments, become imaginary and hence scalarized BH solutions suffer from the gradient instability. We show that such an instability appears irrespective of the structure of the higher-order terms in the GB coupling function and is caused purely due to the existence of the leading quadratic term and the boundary condition that the value of the scalar field vanishes at the spatial infinity.~This indicates that the gradient instability appears at the point in the mass-charge diagram where the scalarized branches bifurcate from the Schwarzschild branch. We also show that scalarized BH solutions realized in a nonlinear scalarization model also suffer from the gradient instability in the even-parity perturbations. Our result also suggests the gradient instability of the exterior solutions of the static and spherically-symmetric scalarized NS solutions induced by the same GB coupling functions.
\end{abstract}

\date{\today}

\maketitle

\section{Introduction}
\label{sec1}

In the context of gravitational theories, scalar fields appear e.g. as a consequence of dimensional reduction of higher-dimensional theories and provide a natural path to extend general relativity  (GR) \cite{Berti:2015itd}. Fundamental issues in cosmology, such the origin of inflation and the late-time acceleration of our Universe, may be considered as an indication of the existence of new gravitational scalar fields beyond GR. In particular, scalar fields have been one of the most popular frameworks to address the accelerated expansions in the early- and late-time stages of our Universe. On the other hand, the suppression of extra scalar forces requires the necessity of a screening mechanism (see e.g., Ref.~\cite{Koyama:2015vza}) which exists beyond the conventional scheme of the scalar-tensor theories. The Horndeski  \cite{Horndeski:1974wa,Deffayet:2009mn,Deffayet:2011gz,Kobayashi:2011nu} and DHOST \cite{Langlois:2015cwa,BenAchour:2016fzp} theories known as the higher-derivative scalar-tensor theories without Ostrogradsky instabilities could implement the suppression mechanisms of extra scalar forces.

Physics of black holes (BHs) and neutron stars (NSs) may also be able to probe the existence of scalar fields in strong gravity regimes~\cite{Berti:2015itd,Barack:2018yly}. BH no-hair theorems hold for scalar-tensor theories with (non)canonical kinetic term, non-negative potential, and non-minimal coupling to the scalar curvature \cite{Israel:1967wq,Carter:1971zc,Ruffini:1971bza,Hawking:1971vc,Chase,Hawking:1972qk,Bekenstein:1995un,Graham:2014mda,Graham:2014ina,Herdeiro:2015waa}, and the shift symmetric subclass of the (beyond-)Horndeski theories with the regular coupling functions of the scalar field $\phi$ and its kinetic term $X:=-\frac{1}{2}g^{\mu\nu}\partial_\mu\phi \partial_\nu\phi$~\cite{Hui:2012qt,Babichev:2017guv,Barausse:2015wia,Capuano:2023yyh}. If the conditions for the no-hair theorem are met, the Schwarzschild or Kerr solutions are the unique vacuum solutions of the gravitational field equations under the given symmetry respectively.

On the other hand, when the scalar field is non-minimally coupled to the Gauss-Bonnet (GB) invariant $\xi(\phi)R_{\rm GB}^2$ in the Lagrangian density where $\xi(\phi)$ is the coupling function and $R_{\rm GB}^2:=R^2-4R^{\alpha\beta}R_{\alpha\beta}+R^{\alpha\beta\mu\nu}R_{\alpha\beta\mu\nu}$ represents the GB invariant, BH solutions with nontrivial profiles  of the scalar field exist because of the violation of the assumptions in the no-hair theorem. Such BH solutions include those for the dilatonic couplings $\xi(\phi)\propto e^{-c\phi}$ \cite{Kanti:1995vq,Alexeev:1996vs,Torii:1996yi,Kanti:1997br,Chen:2006ge,Guo:2008hf,Guo:2008eq,Ohta:2009tb,Ohta:2009pe,Pani:2009wy,Ayzenberg:2014aka,Maselli:2015tta,Kleihaus:2011tg,Kokkotas:2017ymc} and for the linear coupling with the shift symmetry $\xi(\phi)=c' \phi$ \cite{Sotiriou:2013qea,Sotiriou:2014pfa}, with $c$ and $c'$ being constants. For such monotonic couplings, the Schwarzschild or Kerr metrics are no longer solutions.

Models for spontaneous scalarization were first studied for NSs. Even in the simplest model discussed by Damour and Esposito-Far\`ese \cite{Damour:1993hw,Damour:1996ke}, the presence of the scalar field can significantly modify the properties of NSs via the tachyonic instabilities, while satisfying the experimental constraints in the weak-field regimes \cite{Will:2014kxa}. Irrespective of the choice of the equations of state, spontaneous scalarization occurs for the coupling above the certain threshold \cite{Novak:1997hw,Harada:1998ge}, while binary-pulsar observations have put very stringent bounds \cite{Freire:2012mg}. On the other hand, for a long time, there has been no successful model for spontaneous scalarization of BHs. The non-minimal coupling to the Ricci scalar could not trigger the tachyonic instability of the Schwarzschild or Kerr BH, because the Ricci scalar vanishes on such a background. More recently, however, it has been shown that spontaneous scalarization of BHs could also be realized in the presence of the GB coupling $\xi(\phi) R_{\rm GB}^2$ with the $Z_2$ symmetry, $\xi(-\phi)=\xi (\phi)$, such as $\xi(\phi)=\frac{\eta}{8}\phi^2+{\cal O} (\phi^4)$ for the positive quadratic coupling $\eta>0$ \cite{Silva:2017uqg,Doneva:2017bvd,Antoniou:2017acq,Blazquez-Salcedo:2018jnn,Minamitsuji:2018xde,Silva:2018qhn,Cunha:2019dwb,Konoplya:2019fpy,Doneva:2021dqn,East:2021bqk,Doneva:2020nbb,Doneva:2022yqu,Dima:2020yac,Herdeiro:2020wei,Lai:2023gwe} (see also Ref. \cite{Doneva:2022ewd} for a review). Spontaneous scalarization of the Schwarzschild BH requires $\eta>0$, because the GB invariant $R_{\rm GB}^2$ is always positive in the Schwarzschild backgrounds. While BHs with the nontrivial profile of the scalar field for the pure quadratic coupling $\xi(\phi)=\frac{\eta}{8}\phi^2$ \cite{Silva:2017uqg} could not be the endpoint of the tachyonic instability of Schwarzschild BH solutions, nonlinear corrections to the quadratic term $\frac{\eta}{8}\phi^2$ in the coupling function $\xi(\phi)$ could make the static and spherically symmetric scalarized BHs stable against the radial perturbations \cite{Doneva:2017bvd,Blazquez-Salcedo:2018jnn,Minamitsuji:2018xde,Silva:2018qhn}. Then, it was numerically confirmed that such nontrivial BHs could be realized as the endpoint of the instability \cite{Doneva:2021dqn,East:2021bqk}. On the other hand, for highly spinning BHs, spontaneous scalarization could occur for the negative quadratic coupling  $\eta<0$ \cite{Dima:2020yac,Herdeiro:2020wei}, as the GB invariant may change the sign in the highly spinning Kerr backgrounds. Beyond the standard scalarization scheme induced by the linear tachyonic instability, in the Einstein-scalar-GB theory with the $Z_2$-symmetric coupling where the leading term is given by the quartic order term $\phi^4$ or higher ones, scalarization of static and spherically symmetric BHs could be induced nonlinearly because of the existence of the large initial perturbation of the scalar field above a certain threshold value \cite{Doneva:2021tvn,Blazquez-Salcedo:2022omw}. The spinning nonlinearly scalarized BH solutions were constructed recently in Ref.~\cite{Doneva:2022yqu,Lai:2023gwe}. Moreover, in the presence of matter fields, Ref.~\cite{Doneva:2017duq} showed that the same type of the GB coupling function as that in Refs.~\cite{Doneva:2017bvd,Blazquez-Salcedo:2018jnn} could also lead to spontaneous scalarization of NSs, for both signs of the leading quadratic term $\frac{\eta}{8}\phi^2$ in the coupling function $\xi (\phi)$.

On the other hand, it is well-known that the Einstein-scalar-GB theories correspond to a subclass of the Horndeski theories which are known as the most general scalar-tensor theories with the second-order equations of motion. The linear stability analysis of the static and spherically symmetric BH and NS solutions in the Horndeski theories have been performed in the literature \cite{Kobayashi:2012kh,Kobayashi:2014wsa,Kase:2021mix}. These linear stability conditions have been applied to various static and spherically symmetric BH solutions with the nontrivial profile of the scalar field in the Horndeski theories in  Refs. \cite{Minamitsuji:2022mlv,Minamitsuji:2022vbi}. In Ref.~\cite{Minamitsuji:2022mlv}, it has been shown that the static and spherically symmetric BH solutions in reflection-symmetric subclass of shift-symmetric Horndeski theories generically suffer from the ghost or Laplacian instability in the  even-parity perturbations in the vicinity of the BH event horizon, which includes the exact non-asymptotically-flat BHs present for the couplings $G_4 \supset X$~\cite{Rinaldi:2012vy,Anabalon:2013oea,Minamitsuji:2013ura,Cisterna:2014nua} and the exact BH solutions in the models with $G_4 \supset (-X)^{1/2}$ \cite{Babichev:2017guv}. Moreover, in generic non-reflection- and non-shift-symmetric Horndeski theories static and spherically symmetric BH solutions with a non-vanishing constant kinetic term on the horizon $X\neq 0$ inevitably suffer from a ghost or gradient instability \cite{Minamitsuji:2022vbi}. On the other hand, within the perturbative regime, it was shown that only the nontrivial static and spherically symmetric BH solutions which are free from the ghost or gradient instability correspond to those in scalar-tensor theories with the power-law couplings to the GB invariant (with the possible corrections from the regular galileon couplings), which include asymptotically-flat BH solutions in the shift-symmetric theory with the linear coupling to the GB invariant $\phi R_{\rm GB}^2$, which is equivalent to $G_5(X) \propto \ln |X|$~\cite{Sotiriou:2013qea,Sotiriou:2014pfa}. However, scalarized BHs obtained through the coupling to the GB invariant can be realized in the nonperturbative regime with the ${\cal O} (1)$ dimensionless coupling constant normalized by the typical size of the system, e.g., the size of the BH event horizon, and hence are beyond the linear stability analysis in the previous studies.

In this paper, we will investigate the linear stability of the static and spherically symmetric scalarized solutions in the Einstein-scalar-GB theories against the odd- and even-parity perturbations. We will show that the sound speeds for the angular propagations in the even-parity perturbations are imaginary in the limit of  $\ell \to \infty$ with $\ell$ being the angular multipole moments, and hence the static and spherically symmetric scalarized BH solutions will suffer from the gradient instabilities in the angular directions, while they satisfy the other linear stability conditions. We will also show that the appearance of such instabilities is irrespective of the structure of the higher-order terms in the GB coupling functions. Our result will also apply to the exterior solutions of static and spherically symmetric scalarized NSs in the same class of the Einstein-scalar-GB theories, which share the same form of the external solutions as the BH case. Finally, we will show that  the nontrivial BH solutions in a nonlinear scalarization model with the GB coupling function of the form $\xi(\phi)\propto \phi^4+d\phi^6+\cdots$, with $d$ being constant, also suffer from the gradient instability in the angular propagations in the even-parity perturbations. From our analysis, we will expect that the  static and spherically symmetric scalarized BH solutions in the Einstein-scalar-GB theories with any $Z_2$-symmetric coupling functions would generically suffer from the gradient instabilities along the angular propagations in the even-parity perturbations.

The paper is organized as follows: In Sec. \ref{sec2}, we review the Einstein-scalar-GB theory as a subclass of the Horndeski theories. We then discuss the properties of the static and spherically BH solutions in the Einstein-scalar-GB theories including the scalarized BH solutions. In Sec. \ref{sec3}, we review the linear stability conditions of the static and spherically symmetric vacuum solutions with the static scalar field against the odd- and even-parity perturbations in the Horndeski theories. In Sec. \ref{sec4}, we apply the linear stability criteria introduced in Sec. \ref{sec3} to the scalarized BH solutions and the exterior solutions of scalarized NS solutions in the asymptotic limit. In Sec. \ref{sec5}, we also discuss the linear stability of the static and spherically symmetric vacuum solutions which could be realized as the consequence of nonlinear scalarization. The last section \ref{sec6} is devoted to giving a brief summary and conclusion.

\section{Einstein-scalar-Gauss-Bonnet theories and spontaneous scalarization}
\label{sec2}

\subsection{Einstein-scalar-GB theory as the subclass of the Horndeski theories}

We consider the Horndeski theories \cite{Horndeski:1974wa,Deffayet:2009wt,Kobayashi:2011nu}
whose action is composed of the four-independent parts
\begin{eqnarray}
\label{action}
S
&=&
\int d^4x\sqrt{-g}
 {\cal L}
=
\int d^4x\sqrt{-g}
\sum_{i=2}^5 {\cal L}_i,
\end{eqnarray}
with the Lagrangian densities given by 
\begin{eqnarray}
{\cal L}_2
&:=&
G_2(\phi, X),
\nonumber\\
{\cal L}_3
&:=&
-G_3(\phi,X)\Box\phi,
\nonumber\\
{\cal L}_4
&:=&
G_4(\phi, X) R
+G_{4X}(\phi, X)
\left[
\big(\Box\phi\big)^2
-
\left(
\phi^{\alpha\beta}
\phi_{\alpha\beta}
\right)
\right],
\nonumber 
\\
{\cal L}_5
&
:=&
G_5(\phi, X) G_{\mu\nu}\phi^{\mu\nu}
-
\frac{1}{6}G_{5X}(\phi, X)
\left[
(\Box\phi)^3
-3\Box \phi
\left(
\phi^{\alpha\beta}
\phi_{\alpha\beta}
\right)
+
2
\phi_{\alpha}{}^\beta
\phi_\beta{}^\rho
\phi_\rho{}^\alpha
\right],
\end{eqnarray}
where $g_{\mu\nu}$ is the spacetime metric, $R$ and $G_{\mu\nu}$ are the Ricci scalar and Einstein tensor associated with $g_{\mu\nu}$, respectively, $\phi$ is the scalar field, $\phi_\mu=\nabla_\mu \phi$, $\phi_{\mu\nu}=\nabla_\mu\nabla_\nu\phi$, and so on, with $\nabla_\mu$ being the covariant derivative associated with the metric $g_{\mu\nu}$, $X$ represents the canonical kinetic term $X:=-(1/2)g^{\mu \nu}\phi_\mu \phi_\nu$, and $G_{i=2,3,4,5}(\phi, X)$ are free functions of   $\phi$ and $X$.

The Einstein-scalar-GB theory 
\begin{eqnarray}
\label{lagrangian_sgb}
{\cal L}_{\rm sGB}
=
\frac{1}{2\kappa^2}
\left[
R
+\zeta X
+
\xi(\phi)
R_{\rm GB}^2
\right],
\end{eqnarray}
where $\kappa^2(=8\pi G)$ denotes the gravitational constant, $\zeta$ is a constant, and $\xi (\phi)$ is the coupling function  of the scalar field $\phi$ to the GB invariant 
\begin{eqnarray}
R_{\rm GB}^2
:=
R^2
-4R^{\alpha\beta}R_{\alpha\beta}
+ R^{\alpha\beta\mu\nu}R_{\alpha\beta\mu\nu},
\end{eqnarray}
is known as subclass of the Horndeski theories \eqref{action} with the following choice of the coupling functions \cite{Kobayashi:2011nu,Langlois:2022eta}
\begin{eqnarray}
G_2 (\phi, X) 
&=&
\frac{1}{2\kappa^2}
\zeta X
+
\frac{4}{\kappa^2}
\xi^{(4)}(\phi)
X^2
\left(
3-\ln |X|
\right),
\\
G_3 (\phi, X) 
&=&
\frac{2}{\kappa^2}
 \xi^{(3)}(\phi)
X
\left(
7-3\ln |X|
\right),
\\
G_4 (\phi, X) 
&=&
\frac{1}{2\kappa^2}
+
\frac{2\xi^{(2)}(\phi)}{\kappa^2}
X
(2-\ln |X|),
\\
G_5 (\phi, X) 
&=&
-
\frac{2}{\kappa^2}
\xi^{(1)}(\phi)
\ln |X|,
\end{eqnarray}
where $\xi^{(n)}:=\partial^n \xi(\phi)/\partial \phi^n$. We choose $\zeta>0$ so that the scalar field has the correct sign of the kinetic term if the GB coupling vanishes.

\subsection{Static and spherically symmetric scalarized BH solutions}

We assume the static and spherically symmetric spacetime
\begin{eqnarray}
\label{metric}
ds^2
&=&
-f(r) dt^2
+\frac{dr^2}
          {h(r)}
+r^2\gamma_{ab}
     d\theta^a d\theta^b.
\end{eqnarray}
and the static scalar field
\begin{eqnarray}
\label{scalar}
\phi=\phi(r),
\end{eqnarray}
where $f(r)$, $h(r)$, and $\phi(r)$ are the pure functions of the radial coordinate $r$, $\gamma_{ab}$ represents the metric of the unit two-sphere, and the coordinates $\theta^a$ run over the angular directions. Substituting the ansatz \eqref{metric} and \eqref{scalar} into the Lagrangian \eqref{lagrangian_sgb} and varying it with respect to $f$ and $h$, we obtain the equations of motion for $f$ and $h$ which are given by 
\begin{align}
\label{eq1}
&2\left(r+2(1-3h)\xi^{(1)}\phi'\right)h' 
+\frac{1}{2}\left(-4+h(4+\zeta r^2\phi'^2)\right)
-8h(h-1)
\left(
\phi'^2 \xi^{(2)}+\phi''\xi^{(1)}  
\right)
=0,
\\
\label{eq2}
&2\left(r+2(1-3h)\xi^{(1)}\phi'\right)f' 
-\frac{f}{2h}
\left(
4+h (-4+\zeta r^2\phi'^2)
\right)
=0.
\end{align}
Similarly, varying  the Lagrangian \eqref{lagrangian_sgb} with respect to the scalar field $\phi$, we obtain the scalar field equation of motion as 
\begin{align}
\label{eq3}
\phi''
+\frac{1}{2}\left(\frac{4}{r}+\frac{f'}{f}+\frac{h'}{h}\right)\phi'
+\frac{4}{\zeta r^2f}(-1+h)\xi^{(1)} f''(r)
-\frac{2f' \xi^{(1)}}{\zeta r^2 f^2h}
  \left(
    h^2 f'+fh'-f'h-3fhh'
   \right)
=0.
\end{align}
Solving Eq.~\eqref{eq2} with respect to $h^{-1}$, we obtain
\begin{align}
\label{b}
\frac{1}{h}
=\frac{1}{8f}
\left(
4f' (r+2\xi^{(1)}\phi')+f (4-\zeta r^2\phi'^2)
\pm
\sqrt{
-384 f f' \xi^{(1)} \phi'
+
\left(
4f'(r+2\xi^{(1)}\phi')
+f(4-\zeta r^2\phi'^2) 
\right)^2
}
\right).
\end{align}
For the domain of the radial coordinate, we consider $r>r_h$ with $r_h$ ($>0$) being the position of the event (outermost) horizon of the spacetime. We assume that  $f>0$, $f'>0$, $h>0$, $h'>0$ for $r>r_h$, and in the limit of $r\to r_h$ $f\to 0$, $h\to 0$, $\frac{f}{h} \to {\rm constant}$, and $\phi$ and $\phi'$ are regular. Then, in the limit of $r\to r_h$, Eq.~\eqref{eq2} becomes 
\begin{align}
 \left. \frac{f}{h}\right|_{r\to r_h} = 
 \left. \left( r+2\xi^{(1)} \phi' \right) f' \right|_{r\to r_h}, \label{eqn:f/h_athorizon}
\end{align}
which can be obtained only from the $(+)$-branch of Eq.~\eqref{b}, provided that $r+2\xi^{(1)} \phi'|_{r\to r_h}>0$. Thus, in the rest of the paper we only focus on the $(+)$-branch. Obviously, Eq.~\eqref{eqn:f/h_athorizon} implies that
\begin{eqnarray}
\left. \left( r+2\xi^{(1)} \phi' \right) h' \right|_{r\to r_h} = 1. \label{eqn:h'_athorizon}
\end{eqnarray}

Assuming $r+2\xi^{(1)} \phi'|_{r\to r_h}>0$ and substituting the $(+)-$branch of  Eq.~\eqref{b} into Eqs.~\eqref{eq1} and \eqref{eq3}, we obtain a set of the equations which are quasi-linear for $f''(r)$ and $\phi'' (r)$, and, by rearranging them, we obtain the evolution equations for $f(r)$ and $\phi(r)$ with respect to $r$:
\begin{align}
\label{eq_A}
f''(r)= F_f\left[f,f',\phi,\phi' ; r\right],
\qquad 
\phi'' (r)= F_\phi \left[f,f',\phi,\phi'; r\right],
\end{align}
where $F_f$ and $F_\phi$ are the nonlinear combinations of the given variables. After integrating Eq.~\eqref{eq_A} for $f(r)$ and $\phi (r)$ with the given boundary conditions near the event horizon and substituting them into Eq.~\eqref{b}, $h(r)$ can be obtained numerically.

We assume that $\xi(\phi)$ is a $Z_2$-symmetric function across $\phi=0$, $\xi(-\phi)=\xi (\phi)$, and thus satisfies $\xi^{(1)}(0)=\xi^{(3)}(0)=\xi^{(5)}(0)=\cdots=\xi^{(2i+1)}(0)=\cdots=0$ (where $i=3,5,7,\cdots$). Since the GB invariant is topological invariant, we may also set $\xi(0)=0$ without loss of generality, and hence the most general $Z_2$-symmetric coupling function which satisfies our requirement is given by
\begin{align}
\label{general}
\xi(\phi)=
\frac{\eta}{8}
\phi^2
+
\sum_{i=2}^{\infty}
\alpha_{2i}
\phi^{2i},
\end{align}
where $\alpha_{4,6,8,\cdots}$ are the constant coefficients for the higher-order terms. In order to realize the tachyonic instability of the Schwarzschild solution against the radial perturbations, we require $\eta>0$.

Near the horizon $r=r_h$, the solution can be expanded as \cite{Silva:2017uqg,Doneva:2017bvd,Antoniou:2017acq,Minamitsuji:2018xde,Silva:2018qhn} 
\begin{eqnarray}
\label{near_hor1}
f(r)
&=&f_1\left(r-r_h\right)+ {\cal O} \left((r-r_h)^2\right),
\\
\label{near_hor2}
h(r)
&=&
\frac{r_h \zeta}
        {48 \xi^{(1)}(\phi_0)^2}
\left(
r_h^2
-\sqrt{r_h^4-\frac{96\xi^{(1)} (\phi_0)^2}{\zeta}}
\right)
\left(r-r_h\right)+ {\cal O} \left((r-r_h)^2\right),
\\
\label{near_hor3}
\phi(r)
&=&
\phi_0
\left[
1
-
\frac{1}
        {4r_h\phi_0 \xi^{(1)}(\phi_0)}
\left(
r_h^2
-\sqrt{r_h^4-\frac{96\xi^{(1)} (\phi_0)^2}{\zeta}}
\right)
\left(r-r_h\right)
\right]
+ {\cal O} \left((r-r_h)^2\right),
\end{eqnarray}
where $\phi_0$ is the value of the scalar field at the event horizon, $f_1$ is an arbitrary constant representing the time-reparametrization invariance in the static and spherically symmetric spacetimes, and we omit to show the ${\cal O}\left((r-r_h)^2\right)$ terms explicitly. As a consistency check, one can easily confirm that the above assumption $r+2\xi^{(1)} \phi'|_{r\to r_h}>0$ and the relation \eqref{eqn:h'_athorizon} are satisfied, using the above expansions. On the other hand, in the large-distance limit $r\gg r_h$, we obtain the asymptotic solutions 
\begin{align}
\label{bc2}
\frac{f}{f_\infty}&= 1-\frac{2M}{r}+\frac{\zeta M Q^2}{12r^3}
   +\frac{MQ}{6r^4}
\left(
\zeta MQ+24 \xi^{(1)}(\phi_\infty)
\right)
+{\cal O} \left(\frac{1}{r^5}\right),
\nonumber\\
h&=1-\frac{2M}{r}
+\frac{\zeta Q^2}{4r^2}+\frac{\zeta M Q^2}{4r^3}
 +\frac{MQ}{3r^4}
\left(
\zeta MQ+24 \xi^{(1)}(\phi_\infty)
\right)
+{\cal O} \left(\frac{1}{r^5}\right),
\nonumber\\
\phi&=
\phi_\infty
+\frac{Q}{r}+\frac{MQ}{r^2}
+\frac{1}{r^3}\left(\frac{4M^2Q}{3}-\frac{\zeta Q^3}{24}\right)
+\frac{M}{6r^4}
 \left(
 12 M^2 Q
-\zeta Q^3
-\frac{24M \xi^{(1)}(\phi_\infty)}{\zeta}
 \right)
+{\cal O} \left(\frac{1}{r^5}\right),
\end{align}
where $\phi_\infty:=\phi (r\to \infty)$ is the asymptotic value of the scalar field, and $M$ and $Q$ are the mass and the scalar charge, respectively. We note that the constant $f_{\infty}$ ($>0$) also represents the time-reparametrization invariance in the static and spherically symmetric spacetimes, and may be set to unity after the proper rescaling of the time coordinate. From Eq.~\eqref{bc2}, $M$ and $Q$ can be numerically evaluated as 
\begin{align}
\label{m_eva}
M=\frac{r}{2}\left(1-h\right)\Big|_{r\to \infty},
\qquad 
Q=-r^2 \phi'(r)\Big|_{r\to \infty}.
\end{align}
The scalarized solutions connect two values of the scalar field: $\phi_0\neq 0$ near the horizon $r=r_h$ and 
\begin{eqnarray}
\label{bc_infty}
\phi_\infty =0,
\end{eqnarray}
in the large-distance limit.

We focus on the nodeless scalarized solution 
where the scalar field $\phi$ monotonically approaches 0
from a nonzero value on the horizon,
which is known as the  fundamental solution 
and only the stable solution against the radial perturbations \cite{Blazquez-Salcedo:2018jnn,Minamitsuji:2018xde}.
In order to satisfy the boundary condition \eqref{bc_infty},
we require that $\phi$ monotonically approaches $0$ from $\phi_0\neq 0$ on the horizon.
From Eq.~\eqref{near_hor3}, we then have to impose $\phi_0\xi^{(1)} (\phi_0)>0$ near the horizon.
For the $Z_2$-symmetric coupling \eqref{general},
without loss of generality we may assume that 
\begin{align}
\label{firstorder}
\phi_0>0,
\qquad
\xi^{(1)} (\phi_0)>0.
\end{align}
Under the boundary condition \eqref{bc_infty} for the scalarized solutions, with the use of the properties $\xi^{(1)}(0)=\xi^{(3)}(0)=\xi^{(5)}(0)=\cdots=\xi^{(2i+1)}(0)=\cdots=0$ (where $i=3,4,5,\cdots$), Eq. \eqref{bc2} reduces to
\begin{align}
\label{bc20}
\frac{f}{f_\infty}
&= 1-\frac{2M}{r}+\frac{\zeta M Q^2}{12r^3}
   +\frac{\zeta M^2Q^2}{6r^4}
+{\cal O} \left(\frac{1}{r^5}\right),
\nonumber\\
h&=1-\frac{2M}{r}
+\frac{\zeta Q^2}{4r^2}+\frac{\zeta M Q^2}{4r^3}
 +\frac{\zeta M^2Q^2}{3r^4}
+{\cal O} \left(\frac{1}{r^5}\right),
\nonumber\\
\phi&=
\frac{Q}{r}+\frac{MQ}{r^2}
+\frac{1}{r^3}\left(\frac{4M^2Q}{3}-\frac{\zeta Q^3}{24}\right)
+\frac{M}{6r^4}
 \left(
 12 M^2 Q
-\zeta Q^3
 \right)
+{\cal O} \left(\frac{1}{r^5}\right).
\end{align}
As we will see later, in order to check whether one of the stability conditions \eqref{B12con} introduced later is satisfied or not, we need terms higher-order in the power of $r^{-1}$, which are omitted in Eq. \eqref{bc20}. For the stability analysis of the scalarized solution (see Sec. \ref{sec4}) we need to expand the background solution at least up to ${\cal O}(r^{-8})$, and for  the stability analysis of the nonlinearly scalarized solutions (see Sec. \ref{sec5}) we need to do that at least up to ${\cal O}(r^{-12})$. However, since the expression of these higher-order corrections to Eq. \eqref{bc20} are quite involved, we do not show them explicitly here.

In the presence of matter fields, scalarization of NSs could also be realized for the same type of the coupling functions~\cite{Doneva:2017duq}. In contrast to the BH scalarization in the static and spherically symmetric spacetime, since the GB could change the sign in the presence of matter fields, the NS scalarization could occur for both signs of the coupling terms. Assuming the same asymptotic value of the scalar field as Eq. \eqref{bc_infty}, the metric and the scalar field in the exterior solutions for scalarized NSs can also be described by Eq.~\eqref{bc20}.
In this case, the values of the mass and scalar charge, $M$ and $Q$,
are determined via the matching with the interior solutions with matter fields
at the surface of a NS and the use of Eq.~\eqref{m_eva} in the asymptotic limit.

\section{Linear stability against the odd- and even-parity perturbations}
\label{sec3}

In this section, we review the linear stability conditions of the static and spherically symmetric solutions with the static scalar field \eqref{scalar}, which have been obtained in Refs. \cite{Kobayashi:2012kh,Kobayashi:2014wsa,Kase:2021mix} (see also Refs.~\cite{Minamitsuji:2022mlv,Minamitsuji:2022vbi} for applications of these conditions to the concrete BH solutions in the Horndeski theories). As in the case of GR and the conventional scalar-tensor theories, the perturbations about the static and spherically symmetric solutions can be decomposed into the odd- and even-parity perturbations. For the higher-order multipolar modes~$\ell \geq 2$, while the odd-parity perturbations contain just one metric degree of freedom (DOF), i,e., one of the two tensorial polarizations, the even-parity perturbations contain two DOFs, i.e., the other tensorial polarization and the scalar field polarization.

\subsection{Linear stability conditions in the odd-parity perturbations}

The linear stability against the  odd-parity perturbations is ensured under the following three conditions~\cite{Kobayashi:2012kh}:
\begin{eqnarray}
{\cal F} &:=& 2G_4+h\phi'^2G_{5,\phi}-h\phi'^2  \left( \frac12 h' \phi'+h \phi'' \right) 
G_{5,X}>0\,,
\label{cFdef}\\
{\cal G} &:=& 2G_4+2 h\phi'^2G_{4,X}
-h\phi'^2 \left( G_{5,\phi}+{\frac {f' h\phi' G_{5,X}}{2f}} \right)>0 \,,
\label{cGdef}\\
{\cal H} &:=& 2G_4+2 h\phi'^2G_{4,X}-h\phi'^2G_{5,\phi}
-\frac{h^2 \phi'^3 G_{5,X}}{r}
>0\,.
\label{cHdef}
\end{eqnarray}
The squared propagation speeds of the odd-parity perturbations along the radial and angular directions are given, respectively, by 
\begin{eqnarray}
\label{ss1}
c_{r,{\rm odd}}^2=\frac{{\cal G}}{{\cal F}}\,,
\qquad 
c_{\Omega,{\rm odd}}^2=\frac{{\cal G}}{{\cal H}}\,.
\end{eqnarray}
Thus, if all the conditions (\ref{cFdef})-(\ref{cHdef}) are satisfied, all the sound speeds in Eq.~\eqref{ss1} are positive.

\subsection{Linear stability conditions in the even-parity perturbations}

In the even-parity perturbations, the kinetic term of the tensorial polarization has the correct sign for \eqref{cGdef},
and then that for the scalar field has the correct sign~\cite{Kobayashi:2014wsa}, if the following condition is satisfied
\begin{eqnarray}
{\cal K} := 2{\cal P}_1-{\cal F}>0\,,
\label{Kcon}
\end{eqnarray}
with 
\begin{eqnarray}
{\cal P}_1 := \frac{h \mu}{2fr^2 {\cal H}^2} 
\left( 
\frac{fr^4 {\cal H}^4}{\mu^2 h} \right)'\,,\qquad
\mu := \frac{2(\phi' a_1+r\sqrt{fh}{\cal H})}{\sqrt{fh}}\,,
\label{defP1}
\end{eqnarray}
where $a_1$ is given in Appendix~\ref{appA}. In the limit of high frequencies, the conditions for the absence of Laplacian instabilities of the even-parity tensorial polarization $\psi$ and the scalar field polarization $\delta \phi$ along the radial direction are given, respectively, by 
\begin{eqnarray}
c_{r1,{\rm even}}^2 
&=& \frac{\mathcal{G}}{\mathcal{F}}>0\,,
\label{cr1even}
\\
c_{r2,{\rm even}}^2
&=&
\frac{2\phi'[ 4r^2 (fh)^{3/2} {\cal H} c_4 
(2\phi' a_1+r\sqrt{fh}\,{\cal H})
-2a_1^2 f^{3/2} \sqrt{h} 
\phi' {\cal G} 
+( a_1 f'+2 c_2 f ) r^2 fh 
{\cal H}^2]}{f^{5/2} h^{3/2} 
(2{\cal P}_1-{\cal F}) \mu^2}>0
\,,\label{cr2}
\end{eqnarray}
where $c_2$ and $c_4$ are presented in Appendix~\ref{appA}. Since $c_{r1,{\rm even}}^2$ is the same as $c_{r,{\rm odd}}^2$, only the second propagation speed squared~$c_{r2,{\rm even}}^2$ provides an additional stability condition. We note that for the monopole mode $\ell=0$ there is no propagation for the gravitational perturbation, while the scalar-field perturbation~$\delta \phi$ propagates with the same radial velocity as Eq.~(\ref{cr2}). We also note that for the dipole mode $\ell=1$ there is only one gauge DOF for fixing $\delta \phi=0$, under which the gravitational perturbation propagates with the same radial speed squared as Eq.~(\ref{cr2}).

We then turn to the linear stability conditions against the propagation in the angular directions. In the limit of large multipoles~$\ell\gg 1$, the conditions associated with the squared angular propagation speeds in the even-parity perturbations are~\cite{Kase:2021mix,Minamitsuji:2022mlv,Minamitsuji:2022vbi}
\begin{eqnarray}
c_{\Omega \pm}^2=-B_1\pm\sqrt{B_1^2-B_2}>0\,,
\label{cosq}
\end{eqnarray}
where we present the explicit form of $B_1$ and $B_2$ in Appendix~\ref{appA}. These conditions are satisfied if and only if
\begin{eqnarray}
B_1^2 \geq B_2>0 \quad {\rm and} 
\quad B_1<0\,.
\label{B12con}
\end{eqnarray}

\section{Linear stability of scalarized solutions}
\label{sec4}

In this section, we apply the linear stability conditions mentioned in the previous section, Eqs.~\eqref{cFdef}, \eqref{cGdef}, \eqref{cHdef}, \eqref{cr1even}, \eqref{cr2}, and \eqref{B12con} to the static and spherically symmetric scalarized solutions discussed in the literature. For our discussion, we employ the expansion of the metric and scalar field in the large distance region \eqref{bc20} with the boundary condition of the scalar field \eqref{bc_infty}.

\subsection{The quartic-order coupling}

First, we consider the quartic-order coupling function discussed in Ref.~\cite{Minamitsuji:2018xde,Silva:2018qhn} 
\begin{align}
\label{general_quo}
\xi(\phi)=
\frac{\eta}{8}\left( \phi^2+\alpha \phi^4 \right),
\end{align}
where $\eta$ and $\alpha$ are constants. We require that $\eta>0$ and $\alpha<0$, so that the Schwarzschild solution suffers from the tachyonic instability and the scalarized BHs are stable against the radial perturbations \cite{Minamitsuji:2018xde,Silva:2018qhn}, respectively.

In the limit of $r\to\infty$, under the boundary condition \eqref{bc_infty}, the functions $\mathcal{F}$, $\mathcal{G}$, and $\mathcal{H}$ defined in Eqs. \eqref{cFdef}-\eqref{cHdef} can be expanded as
\begin{eqnarray}
\label{mfgh_expand}
\mathcal{F}
&=&
\frac{1}{\kappa^2}
\left(
1-\frac{3Q^2\eta}
           {r^4}
\right)
+
{\cal O}
\left(\frac{1}{r^6}\right),
\quad 
\mathcal{G}
=
\frac{1}{\kappa^2}
\left(
1+\frac{MQ^2\eta}
           {r^5}
\right)
+
{\cal O}
\left(\frac{1}{r^6}\right),
\quad 
\mathcal{H}
=
\frac{1}{\kappa^2}
\left(
1+\frac{Q^2\eta}
           {r^4}
\right)
+
{\cal O}
\left(\frac{1}{r^6}\right),
\end{eqnarray}
which are always positive at the leading order. The sound speeds for the radial and angular propagations in the odd-parity perturbations \eqref{ss1} and that for the radial propagation of the metric perturbations in the even-parity perturbations \eqref{cr1even} coincide with the speed of light at the leading order, with the corrections of ${\cal O}(r^{-4})$. The function $\mathcal{K}$ defined by Eq.~\eqref{Kcon} can be expanded as 
\begin{eqnarray}
\label{mk_expand}
\mathcal{K}=
\frac{\zeta Q^2}{\kappa^2r^2}
\left(
\frac{1}{4}
+\frac{M}{r}
\right)
+
{\cal O} 
\left(\frac{1}{r^4}\right),
\end{eqnarray}
which is also positive for the correct sign of the kinetic term $\zeta>0$. 
The sound speed of the radial perturbations of the scalar field in the even-parity perturbations $c_{r2,{\rm even}}^2$, which can be evaluated via Eq.~\eqref{cr2}, also coincides with the speed of light at the leading order, with the corrections of ${\cal O}(r^{-9})$. Thus, in the large $r$ region the scalarized BHs are linearly stable against all types of propagations in the odd-parity perturbations and against the radial propagations in the even-parity perturbations.

We now check the angular sound speeds of the even-parity perturbations. The functions $B_1$ and $B_2$ can be expanded as 
\begin{eqnarray}
B_1
&=&
-1
+\frac{Q^2\eta}
         {2r^4}
+\frac{Q^2\left(-20M^2+Q^2(24\alpha+\zeta)\right)\eta}
         {24r^6}
+\frac{MQ^2\eta \left(-16M^2\zeta+Q^2\zeta (12\alpha+\zeta)+12\eta\right)}
          {6\zeta r^7}
+
{\cal O} 
\left(\frac{1}{r^8}\right),
\\
B_2
&=&
1
-\frac{Q^2\eta}
         {r^4}
-\frac{Q^2\left(-20M^2+Q^2(24\alpha+\zeta)\right)\eta}
         {12r^6}
+\frac{MQ^2\eta \left(16M^2\zeta-Q^2\zeta (12\alpha+\zeta)+24\eta\right)}
          {3\zeta r^7}
+
{\cal O} 
\left(\frac{1}{r^8}\right),
\end{eqnarray}
leading to
\begin{eqnarray}
\label{b1mb2}
B_1^2-B_2
=
-\frac{12M Q^2\eta^2}{\zeta r^7}
+{\cal O}
\left(
\frac{1}{r^8}
\right),
\end{eqnarray}
which is negative at the leading order and the first condition of Eq.~\eqref{B12con} is not satisfied, since $M>0$ and $\zeta>0$ for the correct sign of the scalar kinetic term. We also note that the above result \eqref{b1mb2} is independent of the sign of $\eta$ and $Q$. In addition, since the leading term in Eq.~\eqref{b1mb2} does not depend on the coefficient of the quartic-order term $\alpha$, we expect that the same leading behavior should be obtained for other nonlinear coupling functions.

\subsection{The exponential coupling}

To confirm our expectation in the previous subsection, we consider the exponential coupling function discussed originally in Ref.~\cite{Doneva:2017bvd}
\begin{align}
\label{general2}
\xi(\phi)
=
\frac{\eta}{8\beta}
\left(
1-e^{-\beta\phi^2}
\right),
\end{align}
which has also been employed in the literature \cite{Doneva:2017bvd,Blazquez-Salcedo:2018jnn,Cunha:2019dwb,Doneva:2021dqn,East:2021bqk,Doneva:2022ewd}. Again, in order to obtain BH solutions which are linearly stable against the radial perturbations, we require that $\eta>0$ and $\beta>0$. In the limit of the spatial infinity, $r\to\infty$, under the boundary condition \eqref{bc_infty}, the functions of $\mathcal{F}$, $\mathcal{G}$, and $\mathcal{H}$ can be expanded as Eq. \eqref{mfgh_expand}, which are always positive. The function $\mathcal{K}$ can be expanded as Eq. \eqref{mk_expand}, which is also positive at the leading order. The sound speeds for the radial and angular propagations in the odd-parity perturbations \eqref{ss1} and that for the radial propagation of the metric perturbations in the even-parity perturbations \eqref{cr1even} coincide with the speed of light at the leading order, with the corrections of ${\cal O}(r^{-4})$. The sound speed of the radial propagation of the scalar field in the even-parity perturbations $c_{r2,{\rm even}}^2$ can be evaluated via Eq.~\eqref{cr2} and also coincides with the speed of light at the leading order, with the corrections of ${\cal O}(r^{-9})$. Thus, in the large $r$ region the scalarized BHs are linearly stable against the odd-parity perturbations and against the radial propagations in the even-parity perturbations.

The functions $B_1$ and $B_2$ can be expanded as 
\begin{eqnarray}
B_1
&=&
-1
+\frac{Q^2\eta}
         {2r^4}
+\frac{Q^2\left(-20M^2+Q^2(-12\beta+\zeta)\right)\eta}
         {24r^6}
+\frac{MQ^2\eta \left(-16M^2\zeta+Q^2\zeta (-6\beta+\zeta)+12\eta\right)}
          {6\zeta r^7}
+
{\cal O} 
\left(\frac{1}{r^8}\right),
\\
B_2
&=&
1
-\frac{Q^2\eta}
         {r^4}
+\frac{Q^2\left(20M^2+Q^2(12\beta-\zeta)\right)\eta}
         {12r^6}
+\frac{MQ^2\eta \left(16M^2\zeta+Q^2\zeta (6\beta-\zeta)+24\eta\right)}
          {3\zeta r^7}
+
{\cal O} 
\left(\frac{1}{r^8}\right),
\end{eqnarray}
which also lead to the same leading behavior of $B_1^2-B_2$ as Eq.~\eqref{b1mb2}, and again we find that the first condition  of Eq.~\eqref{B12con} is not satisfied.

Ref.~\cite{Doneva:2017duq} showed that the same coupling function as Eq. \eqref{general2} could realize scalarization of static and spherically symmetric NSs. In contrast to the case of BH scalarization, since in the static and spherically symmetric spacetimes the GB invariant could change the sign in the presence of matter fields, NS scalarization could occur for both signs of the parameter $\eta$. Since the metric and the scalar field in the exterior solutions for scalarized NSs are also described by Eq.~\eqref{bc20}, our results in the section can also be applied to the case of NS scalarization. As the leading-order term in Eq. \eqref{b1mb2} is irrespective of the sign of $\eta$, the first condition of Eq.~\eqref{B12con} is violated also for NS scalarization with any matter equation of state. Thus, our analysis in this section should exclude both BH and NS scalarization models induced by the GB coupling function of the form \eqref{general2}.

\subsection{More general coupling functions}

Along the same analysis for more general couplings \eqref{general}, we obtain the same leading behavior as Eq. \eqref{b1mb2}. Thus, the instability does not depend on the higher-order structure of the coupling function $\xi (\phi)$. The result that the gradient instability in the even-parity perturbations appears in the angular directions, irrespective of the detailed structure of $\xi(\phi)$, implies that in the large-$\ell$ limit the onset of this instability would take place in the vicinity of the bifurcation point of the scalarized branch from the Schwarzschild branch in the mass-charge diagram, i.e., on the axis of $Q=0$, whose position is also irrespective of the higher-order terms in $\xi(\phi)$.

It would be very interesting if one can obtain universal constraints on the generic coupling functions $\xi(\phi)$ from theoretical considerations. For example, it is known that so-called positivity bounds~\cite{Adams:2006sv} put non-trivial constraints on low-energy effective field theories by assuming the existence of local, causal, unitary, and Lorentz-invariant ultraviolet (UV) completions. However, the positivity bounds in the presence of gravity are rather subtle and still at the stage of development, in particular require some additional assumptions about unknown behaviors of the UV completion of gravity (see e.g. Ref.~\cite{Aoki:2021ckh}). Moreover, even without inclusion of gravity the naive positivity bounds may be violated around Lorentz-violating backgrounds~\cite{Aoki:2021ffc}. Nonetheless, some trials to consider positivity bounds in the context of the Einstein-scalar-GB theories have been made (see e.g. Ref.~\cite{Herrero-Valea:2021dry}). It is certainly worthwhile developing our understanding of positivity bounds in the presence of gravity around Lorentz-violating backgrounds such as BHs.

\section{Linear stability of black holes in a nonlinear scalarization model}
\label{sec5}

We then consider the case of another exponential coupling function discussed originally for nonlinear BH scalarization
in Refs.~\cite{Doneva:2021tvn,Blazquez-Salcedo:2022omw}. 
\begin{align}
\label{general_nl}
\xi(\phi)
=
\frac{\eta_{\rm NL}}{16\beta_{\rm NL}}
\left(
1-e^{-\beta_{\rm NL}\phi^4}
\right),
\end{align}
where $\eta_{\rm NL}$ and $\beta_{\rm NL}>0$ are constants.

In the limit of $r\to\infty$, under the boundary condition \eqref{bc_infty}, the functions $\mathcal{F}$, $\mathcal{G}$, and $\mathcal{H}$ defined in Eqs. \eqref{cFdef}-\eqref{cHdef} can be expanded as
\begin{eqnarray}
\label{mfgh_expand_nl}
&&
\mathcal{F}
=
\frac{1}{\kappa^2}
\left(
1-\frac{5Q^4\eta_{\rm NL}}
           {r^6}
\right)
+
{\cal O}
\left(\frac{1}{r^7}\right),
\quad 
\mathcal{G}
=
\frac{1}{\kappa^2}
\left(
1+\frac{MQ^4\eta_{\rm NL}}
           {r^7}
\right)
+
{\cal O}
\left(\frac{1}{r^8}\right),
\nonumber\\
&&
\mathcal{H}
=
\frac{1}{\kappa^2}
\left(
1+\frac{Q^4\eta_{\rm NL}}
           {r^6}
\right)
+
{\cal O}
\left(\frac{1}{r^6}\right),
\end{eqnarray}
which are always positive at the leading order. The sound speeds for the radial and angular propagations in the odd-parity perturbations \eqref{ss1} and that for the radial propagation of the metric perturbations in the even-parity perturbations \eqref{cr1even} coincide with the speed of light at the leading order. The function $\mathcal{K}$ defined by Eq.~\eqref{Kcon} can be expanded as Eq. \eqref{mk_expand}, which is also positive for $\zeta>0$. The sound speed of the radial propagations of the scalar field in the even-parity perturbations $c_{r2,{\rm even}}^2$, which can be evaluated via Eq.~\eqref{cr2}, also coincides with the speed of light at the leading order. Thus, the scalarized BHs are linearly stable against all types of propagations in the odd-parity perturbations and against the radial propagations in the even-parity perturbations.

The functions $B_1$ and $B_2$ can be expanded as 
\begin{eqnarray}
B_1
&=&
-1
+\frac{Q^4\eta_{\rm NL}}
         {2r^6}
+\frac{M Q^4\eta_{\rm NL}}
         {r^7}
+\frac{M^2 Q^4\eta_{\rm NL}}
          {r^8}
+
\frac{1}{r^9}
\left[
-M^3Q^4 \eta_{\rm NL}
+
\frac{1}{24}
M Q^6 \zeta \eta_{\rm NL}
\right]
\nonumber\\
&&
-
\frac{Q^4\eta_{\rm NL}}
       {480 r^{10}}
\left(
4336 M^4
-
192  M^2 Q^2\zeta
+
Q^4
(240\beta_{\rm NL}+\zeta^2)
\right)
\nonumber\\
&&
+
\frac{1}{r^{11}}
\left[
-\frac{166}{5}M^5Q^4\eta_{\rm NL}
+\frac{89}{40}M^3 Q^6\zeta \eta_{\rm NL}
+\frac{6MQ^6\eta_{\rm NL}^2}{\zeta}
+MQ^8
\left(
-3\beta_{\rm NL}\eta_{\rm NL}
-
\frac{17\zeta^2\eta_{\rm NL}}{640}
\right)
\right]
+
{\cal O} 
\left(\frac{1}{r^{12}}\right),
\\
B_2
&=&
1
-\frac{Q^4\eta_{\rm NL}}
         {r^6}
-\frac{2M Q^4\eta_{\rm NL}}
         {r^7}
-\frac{2M^2 Q^4\eta_{\rm NL}}
          {r^8}
+
\frac{2}{r^9}
\left[
M^3Q^4 \eta_{\rm NL}
-
\frac{1}{24}
M Q^6 \zeta \eta_{\rm NL}
\right]
\nonumber\\
&&
+
\frac{Q^4\eta_{\rm NL}}
       {240 r^{10}}
\left(
4336 M^4
-
192  M^2 Q^2\zeta
+
Q^4
(240\beta_{\rm NL}+\zeta^2)
\right)
\nonumber\\
&&
+
\frac{1}{r^{11}}
\left[
\frac{332}{5}M^5Q^4\eta_{\rm NL}
-\frac{89}{20}M^3 Q^6\zeta \eta_{\rm NL}
+\frac{24MQ^6\eta_{\rm NL}^2}{\zeta}
+MQ^8
\left(
6\beta_{\rm NL}\eta_{\rm NL}
+
\frac{17\zeta^2\eta_{\rm NL}}{320}
\right)
\right]
+
{\cal O} 
\left(\frac{1}{r^{12}}\right),
\end{eqnarray}
leading to
\begin{eqnarray}
\label{b1b2_nl}
B_1^2-B_2
=
-\frac{36MQ^6\eta_{\rm NL}^2}
          {\zeta r^{11}}
+
{\cal O} 
\left(\frac{1}{r^{12}}\right),
\end{eqnarray}
which is negative at the leading order and the first condition  of Eq.~\eqref{B12con} is not satisfied, since $M>0$ and $\zeta>0$ for the correct sign of the scalar kinetic term. We note that the above result \eqref{b1b2_nl} is independent of the sign of $\eta_{\rm NL}$ and $Q$. In addition, the leading term in Eq.~\eqref{b1b2_nl} does not depend on the coefficient of the higher-order terms $\beta_{\rm NL}$. Thus, the nonlinear scalarization model also suffers from the gradient instability along the angular propagations in the even-parity perturbations.

Through the similar analysis, we expect that scalarized BH (and NS) solutions obtained in the Einstein-scalar-GB theory with $Z_2$-symmetric coupling functions where the leading-order term is given by $\phi^6$ or higher-order powers of $\phi$ are also linearly unstable against the angular propagations of the even-parity perturbations. In order to show this, we need to expand the background solutions up to the order of ${\cal O}(r^{-16})$ or even higher order in the power of $r^{-1}$. Since this would require more computation power, we postpone the explicit analysis on such cases for future work. Nevertheless, according to the results so far, it is natural to expect that scalarized BH and NS solutions obtained in the Einstein-scalar-GB theory with any  $Z_2$-symmetric coupling function are also linearly unstable in the case that the asymptotic value of the scalar field is zero, which would exclude models of scalarization induced by the GB coupling at least in the context of the static and spherically symmetric backgrounds.

\section{Summary and conclusions}
\label{sec6}

In this paper, we have investigated the linear stability of the static and spherically-symmetric scalarized BH solutions in the Einstein-scalar-GB theories. Since the Einstein-scalar-GB theories are a subclass of the Horndeski theories, we applied the linear stability conditions for the static and spherically symmetric BH solutions which have been obtained in Refs.~\cite{Kobayashi:2012kh,Kobayashi:2014wsa,Kase:2021mix}. Perturbations about the static and spherically symmetric spacetimes are decomposed into the odd- and even-parity perturbations. While the odd-parity perturbations contain one DOF corresponding to one of the tensorial polarizations, the even-parity perturbations contain two DOFs which correspond to the other tensorial polarization and the polarization of the scalar field. The linear stability conditions are given by Eqs.~\eqref{cFdef}, \eqref{cGdef}, \eqref{cHdef}, \eqref{Kcon}, and \eqref{B12con}.

We have studied three different models for spontaneous scalarization of the static and spherically symmetric solutions, namely,  Eqs~\eqref{general_quo}, \eqref{general2}, and more general \eqref{general}. For all these coupling functions, we showed that while the scalarized BH solutions satisfy the linear stability conditions in the odd-parity perturbations, they do not satisfy the first condition of Eq.~\eqref{B12con}, which means that in the limit of the large angular multipoles $\ell \gg 1$ the sound speeds along the angular propagation in the even-parity perturbations become imaginary and the even-parity perturbations suffer from the gradient instability. Since such a behavior solely depends on the parameter $\eta$ which represents the leading quadratic-order coupling in Eqs~\eqref{general_quo}, \eqref{general2}, and \eqref{general}, our results are independent of the higher-order structure in the GB coupling function $\xi(\phi)$ and the onset of the instabilities in the angular directions arises due to the fact that the leading-order term in the GB coupling function is given by the quadratic order term and the scalar field approaches zero at the spatial infinity. Since in the models of spontaneous scalarization of BHs the onset of the tachyonic instability is governed by the quadratic order term, the instability of the scalarized BHs in the limit of $\ell\to \infty$ arises at the bifurcation point of the scalarized branch from the Schwarzschild branch in the mass-charge diagram. We also argued that the gradient instability of the angular perturbations in the even-parity perturbations could arise in the exterior region of the scalarized NS solutions with the same GB couplings, irrespective of the sign of the quadratic order couplings, as the scalarized BHs and NSs share the same asymptotic form of the metric. Thus, both scalarized BHs and external solutions of scalarized NSs generically suffer from the gradient instability along the angular propagations in the even-parity perturbations.

We now mention the difference from the analysis of Ref.~\cite{Langlois:2022eta}, which studied the odd- and even-parity perturbations of static and spherically symmetric hairy BH solutions in the Einstein-scalar-GB theories and argued that the perturbations are well-behaved (See also Refs. \cite{Minamitsuji:2022mlv,Minamitsuji:2022vbi}). While scalarized solutions discussed in our work are constructed in the nonperturbative regime for which the dimensionless quadratic-order coupling constant becomes of ${\cal O} (1)$, $|\eta|/r_h^2={\cal O} (1)$ (See Eq. \eqref{general}), the authors of Ref.~\cite{Langlois:2022eta} considered the background BH solutions realized as the perturbative deviation from the Schwarzschild solution. We also would like to emphasize that while in scalarized solutions the scalar field is assumed to have the vanishing amplitude $\phi_\infty=0$, Eq.~\eqref{bc_infty}, as the boundary condition at the spatial infinity $r\to \infty$, Ref. \cite{Langlois:2022eta} in general assumed a nonzero amplitude of the scalar field at the spatial infinity. The effective dimensionless coupling constant defined in Eq.~(3.6) in Ref. \cite{Langlois:2022eta}, which was regarded as the expansion parameter for the construction of the background solution, always vanishes for the $Z_2$-symmetric coupling functions \eqref{general} and under the boundary condition \eqref{bc_infty}, $\phi_\infty=0$. Thus, since our work considers the different parameter regimes and the different boundary conditions from those in Ref. \cite{Langlois:2022eta}, we cannot directly compare our results with theirs.

Recently, the authors of Ref. \cite{Kleihaus:2023zzs} argued in the presence of the GB coupling as well as the coupling of the scalar field to the Ricci scalar  that static and spherically symmetric scalarized BHs which are linearly stable against the radial perturbations can be unstable against the perturbations in the $\ell=2$ mode in the even-parity perturbations for above the critical coupling to the Ricci scalar. They also constructed the static and axisymmetric scalarized BH solutions which could be realized as the consequence of the tachyonic instability in the $\ell=2$ sector and clarified the existence of the two new branches of the axisymmetric scalarized BHs, which have prolate and oblate configurations, respectively, and share the same bifurcation points from the radially stable branch. With the results in Ref. \cite{Kleihaus:2023zzs} that scalarized BHs stable against the $\ell=0$ perturbations could be unstable against the $\ell=2$ perturbations, as well as the fact that after the decomposition into the angular multipole modes the $\ell$-dependence in the equation of motion for the scalar field perturbation in the decoupling limit appears only in the form of $c_{\Omega}^2\ell (\ell+1)$ with $c_{\Omega}^2$ being the angular sound speed squared, we can speculate that  such an instability occurs only in the case of the imaginary angular sound speed $c_{\Omega}^2<0$ and appears more efficiently for higher multipole modes $\ell>2$. We thus expect that the same type of instabilities should exist for arbitrary larger values of multipole moments $\ell \geq 2$, and then the bifurcation point of the scalarized branches with the $\ell$-th order deformation would be shifted to the direction of the smaller scalar charge $Q$ in the mass-charge diagram. In the limit of $\ell \to \infty$, having speculated from our results in this paper, the bifurcation points of the new scalarized branches with the $\ell$-th order deformation approaches the bifurcation point of the radially stable scalarized branches from the Schwarzschild branch in the mass-charge diagram, which is on the axis of the vanishing scalar charge $Q=0$. As the consequence, all the scalarized BHs are unstable against the small angular deformation on the event horizon. A further inspection of such  deformed scalarized BHs will be left for future work.

We also analyzed the linear stability of BH solutions in a nonlinear scalarization model, whose coupling function is given by Eq. \eqref{general_nl}. We also showed that in this model BHs with the vanishing asymptotic value of the scalar field do not satisfy the first condition of Eq.~\eqref{B12con}, and suffer from the gradient instability of even-parity perturbations in the angular directions. Since the leading-order term of $B_1^2-B_2$ does not depend on the parameter for the higher-order terms $\beta_{\rm NL}$, we also expect that such an instability is due to the existence of the leading quartic-order coupling as well as the vanishing asymptotic amplitude of the scalar field at the spatial infinity. From the above results, we also expect that the static and spherically symmetric BH solutions with the vanishing asymptotic values of the scalar field realized with any $Z_2$-symmetric GB coupling function suffer from the gradient instability of the angular propagations in the even-parity perturbations. On the other hand, as demonstrated for the linear coupling model with the shift symmetry in Appendix~\ref{appB}, BHs with the nontrivial scalar field in the Einstein-scalar-GB theories with non-$Z_2$-symmetric coupling functions would be linearly stable for all types of perturbations, at least in the large $r$ region. In summary, our results would exclude all the static and spherically symmetric BH and NS solutions realized in both the spontaneous and nonlinear scalarization models with the $Z_2$-symmetric coupling functions.

\section*{ACKNOWLEDGMENTS}
M.M.~was supported by the Portuguese national fund through the Funda\c{c}\~{a}o para a Ci\^encia e a Tecnologia in the scope of the framework of the Decree-Law 57/2016 of August 29, changed by Law 57/2017 of July 19, and the Centro de Astrof\'{\i}sica e Gravita\c c\~ao through the Project~No.~UIDB/00099/2020. The work of SM was supported in part by World Premier International Research Center Initiative, MEXT, Japan.

\appendix

\section{Coefficients associated with perturbations}
\label{appA}

The quantities~$a_1$, $c_2$, and $c_4$ in Eqs.~(\ref{defP1}) 
and (\ref{cr2}) are given by 
\begin{eqnarray}
a_1&=&\sqrt{fh} \left\{  \left[ G_{4,\phi}+\frac12 h ( G_{3,X}-2 G_{4,\phi X} ) \phi'^2 \right] r^2
+2 h \phi' \left[ G_{4,X}-G_{5,\phi}-\frac12h ( 2 G_{4,XX}-G_{5,\phi X} ) \phi'^2 \right] r
\right.
\notag\\
&&
\left.
+\frac12 G_{5,XX} h^3\phi'^4
-\frac12 G_{5,X} h ( 3 h-1 ) \phi'^2 
\right\}\,,\\
c_2&=&\sqrt{fh} \left\{  \left[  
\frac{1}{2f}\left( -\frac12 h ( 3 G_{3,X}-8 G_{4,\phi X} ) \phi'^2
+\frac12 h^2 ( G_{3,XX}-2 G_{4,\phi XX} ) \phi'^4
-G_{4,\phi} \right) r^2
\right.\right.
\notag\\
&&
\left.\left.
-{\frac {h\phi'}{f}} \left( 
\frac12 {h^2 ( 2 G_{4,XXX}-G_{5,\phi XX} ) \phi'^4}
-\frac12 {h ( 12 G_{4,XX}-7 G_{5,\phi X} ) \phi'^2}
+3 ( G_{4,X}-G_{5,\phi} ) \right) r
\right.\right.
\notag\\
&&
\left.\left.
+\frac{h\phi'^2}{4f}\left(
G_{5,XXX} h^3\phi'^4
- G_{5,XX} h ( 10 h-1 ) \phi'^2
+3 G_{5,X}  ( 5 h-1 ) 
\right) \right] f'
\right.
\notag\\
&&
\left.
+\phi' \left[ \frac12G_{2,X}-G_{3,\phi}
-\frac12 h ( G_{2,XX}-G_{3,\phi X} ) \phi'^2 \right] r^2
\right.
\notag\\
&&
\left.
+ 2\left[ -\frac12h ( 3 G_{3,X}-8 G_{4,\phi X} ) \phi'^2
+\frac12h^2 ( G_{3,XX}-2 G_{4,\phi XX} ) \phi'^4
-G_{4,\phi} \right] r
-\frac12 h^3 ( 2 G_{4,XXX}-G_{5,\phi XX} ) \phi'^5
\right.
\notag\\
&&
\left.
+\frac12 h \left[ 2\left(6 h-1\right) G_{4,XX}+\left(1-7 h\right)G_{5,\phi X} \right] \phi'^3
- ( 3 h-1 )  ( G_{4,X}-G_{5,\phi} ) \phi' \right\} \,,\\
c_4&=&\frac14 \frac {\sqrt {f}}{\sqrt {h}} 
\left\{ \frac {h\phi'}{f} \left[ 
2 G_{4,X}-2 G_{5,\phi}
-h ( 2 G_{4,XX}-G_{5,\phi X} ) \phi'^2
-{\frac {h\phi'  ( 3 G_{5,X}-G_{5,XX} \phi'^2h ) }{r}} \right]f'
\right.
\notag\\
&&
\left.
+4 G_{4,\phi}
+2 h ( G_{3,X}-2 G_{4,\phi X} ) \phi'^2
+{\frac {4 h ( G_{4,X}-G_{5,\phi} ) \phi'-2 h^2 ( 2 G_{4,XX}-G_{5,\phi X} ) \phi'^3}{r}} \right\} \,.
\end{eqnarray}
The quantities~$B_1$ and $B_2$ in 
Eq.~(\ref{cosq}) are 
\begin{eqnarray}
&&
B_1=
\frac {r^3\sqrt {f h} {\cal H} [ 4 h ( \phi' a_1+r\sqrt{fh} {\cal H}) 
\beta_1+\beta_2-4 \phi' a_1 \beta_3] 
-2 fh {\cal G}  [ r \sqrt{fh}( 2 {\cal P}_1-{\cal F}){\cal H}  
( 2\phi' a_1+r\sqrt{fh} {\cal H} )+2\phi'^2a_1^{2}{\cal P}_1 ] }
{4f h ( 2 {\cal P}_1-{\cal F} ){\cal H}
(\phi' a_1+r\sqrt{fh} {\cal H})^2}\,,
\label{B1def}
\notag\\\\
&&
B_2=
-r^2{\frac {r^2h \beta_1 [ 2 fh {\cal F} {\cal G} ( \phi' a_1+r \sqrt{fh}{\cal H} ) 
+r^2\beta_2 ] -{r}^{4}\beta_2 \beta_3
-fh{\cal F} {\cal G}  ( \phi' fh {\cal F} {\cal G}a_1 +2 r^3 \sqrt{fh} {\cal H} \beta_3 ) }
{ fh\phi' a_1 ( 2 {\cal P}_1-{\cal F} ){\cal F}  ( \phi' a_1+r \sqrt{fh}{\cal H} ) ^{2}}}\,,
\label{B2def}
\end{eqnarray}
where
\begin{eqnarray}
\beta_1&=&\frac12 \phi'^2 \sqrt{fh} {\cal H}e_4 
-\phi' \left(\sqrt{fh}{\cal H} \right)' c_4 
+ \frac{\sqrt{fh}}{2}\left[ \left( {\frac {f'}{f}}+{\frac {h'}{h}}-\frac{2}{r} \right) {\cal H}
+{\frac {2{\cal F}}{r}} \right] \phi' c_4+{\frac {f{\cal F} {\cal G}}{2r^2}}\,,\\
\beta_2&=& \left[ \frac{\sqrt{fh}{\cal F}}{r^2} \left( 2 hr\phi'^2c_4
+\frac{r \phi' f' \sqrt{h}}{2\sqrt{f}}{\cal H}-\phi' \sqrt{fh}{\cal G} \right)
-\frac{\phi' fh {\cal G}{\cal H}}{r} \left( \frac{{\cal G}'}{{\cal G}}
-\frac{{\cal H}'}{{\cal H}}+\frac{f'}{2f}-\frac{1}{r} \right) \right]a_1 
-\frac{2}{r} (fh)^{3/2}{\cal F}{\cal G}{\cal H}\,, \qquad\,\,\\
\beta_3&=& \frac{\sqrt{fh}{\cal H}}{2}\phi'  
\left( hc_4'+\frac12 h' c_4-\frac{d_3}{2} \right) 
-\frac{\sqrt{fh}}{2} \left( \frac{\cal H}{r}+{\cal H}' \right) 
\left( 2 h \phi'c_4+\frac{\sqrt{fh}{\cal G}}{2r}
+\frac{f'\sqrt{h}{\cal H}}{4\sqrt{f}} \right)\notag\\
&&
+{\frac {\sqrt {fh}{\cal F}}{4r} \left(  2 h \phi'c_4
+\frac{3\sqrt{fh}{\cal G}}{r}
+\frac{f'\sqrt{h}{\cal H}}{2\sqrt{f}}
 \right) }\,,
\end{eqnarray}
with
\begin{eqnarray}
e_4 &=&{\frac {1}{\phi'}}c_4'-{\frac {f'}{4f h \phi'^2}} 
\left( \sqrt{fh} {\cal H} \right)'
-{\frac {\sqrt {f}}{2\phi'^2\sqrt {h}r}}{\cal G}'
+{\frac {1}{h\phi' r^2} \left( {\frac {\phi''}{\phi'}}+\frac12 {\frac {h'}{h}} \right) }a_1
\notag\\
&&
+{\frac {\sqrt{f}}{8\sqrt{h}\phi'^2} 
\left[ {\frac { ( f' r-6 f ) f'}{f^2r}}
+\frac {h' ( f' r+4 f ) }{fhr}
-{\frac {4f ( 2 \phi'' h+h' \phi')}
{\phi' h^2r ( f' r-2 f ) }} \right] }{\cal H}
+{\frac {h'}{2h\phi'}}c_4
-\frac{f'r-2f}{4\sqrt{fh}r\phi'}
\frac{\partial {\cal H}}
{\partial \phi}
\notag\\
&&
+{\frac {f' hr-f}{2r^2\sqrt {f}{h}^{3/2}\phi'^2}}{\cal F}
+{\frac {\sqrt {f}}{2r\phi'^2{h}^{3/2}} 
\left[ {\frac {f ( 2 \phi'' h+h' \phi' ) }{h\phi'  ( f' r-2 f ) }}
+{\frac {2 f-f' hr}{2fr}} \right] }{\cal G}
\,, \label{e4}\\
d_3&=&
-{\frac {1}{r^2} \left( {\frac {2\phi''}{\phi'}}+{\frac {h'}{h}} \right) }a_1
+{\frac {f^{3/2}h^{1/2}}{ ( f' r-2 f ) \phi'} \left( 
{\frac {2\phi''}{h\phi' r}}
+ {\frac {{f'}^{2}}{f^2}}
- {\frac {f' h'}{fh}}
-{\frac {2f'}{fr}}
+{\frac {2h'}{hr}}
+ {\frac {h'}{h^2r}} \right) }{\cal H}
\notag\\
&&
+\frac{f'r-2f}{2r} \sqrt{\frac{h}{f}} 
\frac{\partial{\cal H}}{\partial \phi}
+{\frac {\sqrt {f}}{\phi' \sqrt {h}r^2}}{\cal F}
-{\frac {{f}^{3/2}}{\sqrt {h} ( f' r-2 f ) \phi'} 
\left( {\frac {f'}{fr}}+{\frac {2\phi''}{\phi' r}}+{\frac {h'}{hr}}-\frac{2}{r^2} \right) }{\cal G}
\,.
\end{eqnarray}

\section{Linear stability of BHs in the shift -symmetric scalar-Gauss-Bonnet theory}
\label{appB}

For reference, we consider the linear stability of BH solutions with the nontrivial scalar field in the shift -symmetric scalar-GB theory with the linear coupling 
\begin{align}
\label{linear}
\xi(\phi)=\gamma \phi,
\end{align}
which were discussed in Ref.~\cite{Sotiriou:2013qea,Sotiriou:2014pfa}, where $\gamma$ is constant. For such a linear coupling, the expansion of the background solutions \eqref{bc2} in the large-$r$ limit
reduces to 
\begin{align}
\label{bclinear}
\frac{f}{f_\infty}
&= 1-\frac{2M}{r}+\frac{\zeta M Q^2}{12r^3}
   +\frac{MQ}{6r^4}
\left(
\zeta MQ+24 \gamma
\right)
+{\cal O} \left(\frac{1}{r^5}\right),
\nonumber\\
h&=1-\frac{2M}{r}
+\frac{\zeta Q^2}{4r^2}+\frac{\zeta M Q^2}{4r^3}
 +\frac{MQ}{3r^4}
\left(
\zeta MQ+24 \gamma
\right)
+{\cal O} \left(\frac{1}{r^5}\right),
\nonumber\\
\phi&=
\phi_\infty
+\frac{Q}{r}+\frac{MQ}{r^2}
+\frac{1}{r^3}\left(\frac{4M^2Q}{3}-\frac{\zeta Q^3}{24}\right)
+\frac{M}{6r^4}
 \left(
 12 M^2 Q
-\zeta Q^3
-\frac{24M \gamma}{\zeta}
 \right)
+{\cal O} \left(\frac{1}{r^5}\right),
\end{align}
where the asymptotic value of the scalar field $\phi_\infty$ has no physical meaning due to the shift symmetry.

Using the solution in the limit of the spatial infinity $r\to\infty$, Eq. \eqref{bclinear}, the functions $\mathcal{F}$, $\mathcal{G}$, and $\mathcal{H}$ defined in Eqs. \eqref{cFdef}-\eqref{cHdef} can be expanded as
\begin{eqnarray}
\label{mf_expand_shift}
\mathcal{F}
&=&
\frac{1}{\kappa^2}
\left(
1-\frac{8Q\gamma}
           {r^3}
\right)
+
{\cal O}
\left(\frac{1}{r^4}\right),
\quad 
\mathcal{G}
=
\frac{1}{\kappa^2}
\left(
1+\frac{4MQ\gamma}
           {r^4}
\right)
+
{\cal O}
\left(\frac{1}{r^5}\right),
\quad 
\mathcal{H}
=
\frac{1}{\kappa^2}
\left(
1+\frac{4Q\gamma}
           {r^3}
\right)
+
{\cal O}
\left(\frac{1}{r^5}\right),
\end{eqnarray}
which are always positive at the leading order. Thus, BH solutions are linearly stable against the odd-parity perturbations.

The function $\mathcal{K}$ defined by Eq.~\eqref{Kcon} can be expanded as Eq.~\eqref{mk_expand}, which is also positive. We also show that the sound speeds along the radial propagation in the even parity perturbations \eqref{cr1even}, and \eqref{cr2} coincide with the speed of light at the leading order, with the corrections of ${\cal O} (r^{-3})$. Regarding the angular propagations in the even-parity perturbations, we obtain the leading behavior of the functions $B_1$ and $B_2$
\begin{eqnarray}
B_1
&=&
-1
+\frac{2Q\gamma}
         {r^3}
-\frac{2MQ\gamma}
         {r^4}
+\frac{-16M^2Q\gamma+Q^3\gamma \zeta}
          {r^5}
+
\frac{2\gamma}{3r^6}
\left[
-12M^3Q-12Q^2\gamma
-\frac{480 M^2\gamma}{\zeta}
+
M Q^3\zeta
\right]
+
{\cal O} 
\left(\frac{1}{r^7}\right),
\\
B_2
&=&
1
-
\frac{4Q\gamma}
         {r^3}
+
\frac{4MQ\gamma}
         {r^4}
+\frac{16M^2Q\gamma-Q^3\gamma \zeta}
          {2r^5}
+
\frac{4\gamma}{3r^6}
\left[
12M^3Q
+12Q^2\gamma
+\frac{48 M^2\gamma}{\zeta}
-
M Q^3\zeta
\right]
+
{\cal O} 
\left(\frac{1}{r^7}\right),
\end{eqnarray}
and hence
\begin{eqnarray}
\label{b1mb2_2}
B_1^2-B_2
=
\frac{4\gamma^2 \left(144M^2+\zeta Q^2\right)}
        {\zeta r^6}
+{\cal O}
\left(
\frac{1}{r^7}
\right),
\end{eqnarray}
which is positive at the leading order for the correct sign of the scalar kinetic term $\zeta>0$. Thus, in contrast to the case of the $Z_2$-symmetric GB couplings discussed in the main text, the two conditions of Eq.~\eqref{B12con} are satisfied. The results obtained in this Appendix are consistent with Ref.~\cite{Minamitsuji:2022mlv}.

\bibliography{refs}
\end{document}